\documentclass[12pt]{article}
\newcommand{\be}{\begin{equation}}
\newcommand{\ee}{\end{equation}}
\newcommand{\ba}{\begin{eqnarray}}
\newcommand{\ea}{\end{eqnarray}}

\usepackage{psfig}

\begin{document}
\title{Double electron capture in $^{156}$Dy, $^{162}$Er and $^{168}$Yb}

\author{
Victoria E. Cer\'on\,$^a$ and Jorge G. Hirsch\,$^{b}$. \\
{\small\it $^a$ Departamento de F\'{\i}sica, Centro de
Investigaci\'on y Estudios Avanzados del IPN,} \\
{\small\it Ap. Postal 14-740, 07000 M\'exico D.F., M\'exico}  \\
{\small\it $^b$ Instituto de Ciencias Nucleares, UNAM,} \\
{\small\it Circuito Exterior C.U., Ap. Postal 70-543, 04510 M\'exico
D. F., M\'exico    }  \\
}
\maketitle

\begin{abstract}
The double electron capture half-lives of $^{156}$Dy, $^{162}$Er and
$^{168}$Yb are evaluated using the pseudo SU(3) model, which describes
ground and excited bands as well as their B(E2) and B(M1) transition
strengths in remarkable agreement with experiment. The best candidate for
experimental detection is the decay $^{156}$Dy $\rightarrow ^{156}$Gd,
with $\tau^{1/2} (0^+_{gs} \rightarrow 0^+_{gs}) = 2.74 \times 10^{22}$ 
yrs and $\tau^{1/2} (0^+_{gs} \rightarrow 0^+_1) = 8.31 \times 10^{24}$
yrs.

~~

\end{abstract}

The double beta decay and its relationship with weak-interactions and
neutrino physics have been intensively studied in recent years.  Measuring
$\beta^-\beta^-\,_{2\nu}$ lifetimes of the order $10^{19}-10^{22}$ yrs and
establishing the limit $\langle m_{\nu_e} \rangle \leq 1$eV for the
Majorana mass of the neutrino are major experimental achievements
\cite{Suh98}. 

The double positron emitting decay ($\beta^+\beta^+\,_{2\nu}$) and the
accompanying electron capture (EC) processes have been more elusive. Many
candidates have low natural abundances or Q-values \cite{Boe92}.
While the relevance of $\beta^+\beta^+\,_{0\nu}$  as a lepton violating
process has been known for many years \cite{Ver83}, only recently 
nuclear matrix elements were calculated in the context of the pn-QRPA
\cite{Sta91}. Searches for these decays, both with and without
neutrino emission, have been performed in
$^{54}$Fe, $^{78}$Kr, $^{92}$Mo, $^{106}$Cd and $^{130,132}$Ba
\cite{Sae94}.
 
Double electron capture processes (ECEC$_{2\nu}$) have larger Q-values,
but the decay to the ground state of the final nuclei ($0^+_{gs}
\rightarrow 0^+_{gs}$) has only X-rays emitted, making its detection
difficult \cite{Suh98}. The ECEC$_{2\nu}$ decay to an excited state in the
final nuclei ($0^+_{gs} \rightarrow 0^+_{1}$) could be detected through
the two characteristic gamma rays emitted by the final nuclei \cite{Bar94}

In the present letter we study the ECEC$_{2\nu}$ of $^{156}$Dy, $^{162}$Er
and $^{168}$Yb using the pseudo SU(3) model. The motivation for this
study is twofold: i) The first two nuclei have been mentioned as possible
candidates for experimental detection, with half-lives calculated using
rough estimates for the nuclear matrix elements \cite{Bar94}. ii) The QRPA
has undesirable features which makes its predictions unreliable 
\cite{Vog98}, while shell model calculations provide more
reliable nuclear matrix elements for light and medium mass nuclei
\cite{Vog98,Nak96}. The pseudo SU(3)
shell model describes many features of heavy deformed nuclei, as ground
and excited bands, B(E2) and B(M1) transition strengths in remarkable
agreement with experiment  using a symmetry truncated shell model theory
with Hamiltonians which includes single particle energies,
quadrupole-quadrupole and pairing interactions. The
same formalism has proved its effectiveness for even-even \cite{Beu98}
and odd-even \cite{Var99} nuclei.

As mentioned above the $\beta^+\beta^+\,_{2\nu}$ decay can occur in three
different ways:
\begin{enumerate}
\item The  $\beta^+\beta^+\,_{2\nu}$ proper 
\begin{equation}
(A, Z+2) \longrightarrow (A, Z) + 2e^{+} + 2\nu
\end{equation}
is easy to detect through the annihilation of the two positrons, but
strongly suppressed by the low Q-value and the Coulomb repulsion between
the positrons and the atomic nuclei.

\item The  $\beta^+EC\,_{2\nu}$, which captures one bound electron $e^-_b$
\begin{equation}
e^{-}_b +(A, Z+2) \longrightarrow (A, Z) + e^{+} + 2\nu
\end{equation}
shares with first one the  hindering factors.

\item The ECEC$_{2\nu}$ double electron capture 
\begin{equation}
2e^{-}_b +(A, Z+2) \longrightarrow (A, Z) + 2\nu 
\end{equation}
has the largest Q-values, no Coulomb suppression but is very difficult to
detect, because only two X rays are emitted together with the neutrinos.
\end{enumerate}

The double electron capture decay to excited states in the final nuclei
\begin{eqnarray}
(A,Z+2) + 2e^{-}_b \longrightarrow &(A, Z)^{*} + 2\nu
~~~~~~~~~~~~~~~  \\
& ~~~~^{\mid}\!\!\!\longrightarrow (A, Z) + 2 \gamma
\end{eqnarray}
has been proposed as a good candidate to be measured \cite{Ver83,Bar94}.
The two gammas are far easier to detect than the X rays. 
A sensitivity close to $\sim 10^{22}$yr has been estimated for this type
of experiments \cite{Bar94}.

 The decay scheme for $^{156}$Dy is shown in Fig. 1. It can proceed
directly to the ground state of $^{156}$Gd,  or to two excited $0^+$
states, which decay mainly to the first $2^+$ state emitting 960.55 keV or
1079.24 keV gammas, followed by a gamma with energy 88.97 keV from the
decay of the $2^+$ state \cite{nndc}. The two gammas in coincidence are
the signature of the double electron capture process.

In the pseudo $SU(3)$ shell model coupling scheme \cite{Rat73} normal
parity orbitals $(\eta ,l,j)$ are identified with orbitals of a harmonic 
oscillator of one quanta
less $\tilde \eta = \eta-1$.  This set of orbitals with $\tilde j
= j = \tilde  l + \tilde s$,  pseudo spin $\tilde s =1/2$ and  pseudo
orbital angular momentum  $\tilde l$
define the so called pseudo space. The orbitals with
$j = \tilde l \pm 1/2$ are nearly degenerate. For configurations  
of identical particles occupying a single j orbital of abnormal parity
a convenient characterization of states is made by means of the seniority
coupling scheme.
      
The many particle states of $n_\alpha$ nucleons in a given shell
$\eta_\alpha$,
$\alpha = \nu $ or $\pi$, can be defined by the totally antisymmetric
irreducible representations
$\{ 1^{n^N_\alpha}\} $ and $\{1^{n^A_\alpha}\}$ of unitary groups.
The dimensions of the normal $(N)$ parity space is
$\Omega^N_\alpha = (\tilde\eta_\alpha + 1) (\tilde\eta_\alpha +2)$ and   
that of the unique $(A)$ space is $\Omega^A_\alpha =
2\eta_\alpha +4$ with the constraint
$n_\alpha = n^A_\alpha  + n^N_\alpha$.
Proton and neutron states are coupled to angular momentum $J^N$ and $J^A$
in both the normal and unique parity sectors, respectively. The
wave function of the many-particle state  with angular
momentum $J$ and projection $M$ is expressed as a direct product of the
normal  and unique parity ones, as:

\begin{equation}
|J M > = \sum\limits_{J^N J^A} [|J^N> \otimes |J^A>]^J_M
\end{equation}
We are interested in describing the low-lying energy states.
 In the normal subspace
only pseudo spin zero configurations are taken into account.
Additionally in the abnormal parity space only seniority zero
configurations
 are taken into account.  This simplification implies that
 $J^A_\pi = J^A_\nu = 0$. This is a very
strong assumption quite useful in order to simplify the calculations.
It is possible to describe the intruder states using an SU(3) formalism in
the {\em real} space including S = 0 and 1 states \cite{Var98}.
Calculations performed in this scheme involve a more sophisticated
formalism which we are just starting to explore. 

The double beta decay, when described in the pseudo SU(3) scheme, is
strongly dependent on the occupation numbers for protons and neutrons in
the normal and abnormal parity states $n^N_\pi, n^N_\nu, n^A_\pi,
n^A_\nu$ \cite{Cas94}.
These numbers are determined filling the Nilsson levels from below, as
discussed in \cite{Cas94}.
In particular the $\beta^+ \beta^+$ ~decay is allowed only if
they fulfil the following relationships
 
\begin{equation}
\begin{array}{l}
n^A_{\pi ,f} = n^A_{\pi ,i} - 2~~,
\hspace{1cm}n^A_{\nu ,f} = n^A_{\nu ,i}~~, \\
n^N_{\pi ,f} = n^N_{\pi ,i}~~ ,
\hspace{1.7cm} n^N_{\nu ,f} = n^N_{\nu ,i} + 2 ~~.\label{num}
\end{array}
\end{equation}

\noindent
If these relations are not satisfied the $\beta^+ \beta^+$ decay becomes
forbidden.
This is the first selection rule the pseudo SU(3) formalism imposes on the
double beta decay, similar to that found for $\beta^-\beta^-$ processes
\cite{Cas94}.

In $^{156}$Gd there is one dominant component in the ground state wave
function\cite{Beu98}. Assuming a small deformation to satisfy Eq.
(\ref{num}) the ground state of this nuclei can be described as

\begin{eqnarray}
|^{156}\hbox{Gd}, 0^+\rangle  \approx&|&\{{2^{5}}\}_{\pi} \,(10,4)_{\pi}
,\,\{{2^{4}}\}_{\nu}
\,(18,4)_{\nu};\,(28,8)1 \, L=M=0>_{N}  \nonumber \\
&&  \nonumber \\
&|&(h_{11/2})_{\pi}^{6} \, J_{\pi}^{A} = 0, \;(i_{13/2})_{\nu}^{4} \,
J_{\nu}^{A} = 0 >_{A} .  \label{sdg}
\end{eqnarray}

In a similar way, the ground state of  $^{156}$Dy will be dominated by

\begin{eqnarray}
|^{156}\hbox{Dy}, 0^+\rangle  \approx &|& \{{2^{5}}\}_{\pi} \,(10,4)_{\pi}
;\,\{{2^{3}}\}_{\nu} \,(18,0)_{\nu};\,(28,4)1 \, L=M=0>_{N}  \nonumber \\
&&  \nonumber \\
&|&(h_{11/2 })_{\pi}^{6} \, J_{\pi}^{A} = 0,\;(i_{13/2})_{\nu}^{2} \,
J_{\nu}^{A} = 0 >_{A} .\label{sdy}
\end{eqnarray}

The inverse half life of the two neutrino mode of the $ECEC_{2\nu}$-decay
can be expressed in the form \cite{Doi85}

\begin{equation}
        \left[\tau^{1/2}_{2\nu}(0^+ \rightarrow 0^+_\sigma)\right]^{-1} =
      G_{2\nu}(\sigma) \ | \ M_{2\nu}(\sigma) \ |^2 \ \ .
\end{equation}

\noindent
where $\sigma$= g.s., 1 or 2 labels the final state $0^+_\sigma$,  
$G_{2\nu}(\sigma)$ are kinematical factors which
depend on $E_{\sigma} = {1 \over 2} [{\it Q}_{\beta  \beta}-
E(0^+_\sigma )]$ which is half of the total energy released when 
the electron binding energies are neglected. The nuclear matrix element
is evaluated using the pseudo SU(3) formalism \cite{Hir95}.
For the $^{156}$Gd $\rightarrow$ $^{156}$Dy case it can be written as

\begin{equation}
\begin{array}{ll}
M_{2\nu}^{GT}(\sigma)~ =~a~b(n^A_\pi)  ~{\cal
E}_{\sigma}^{-1}\\
\hspace{1cm}\sum\limits_{(\lambda_0 \mu_0 ) K_0}
<(\tilde\eta 0)1 \tilde l,(\tilde\eta 0)1 \tilde l \| (\lambda_0 \mu_0
) 0 0>_1 \sum\limits_\rho
<(18,0)1~0,(\lambda_0 \mu_0 )K_0 J\|(18,4 )
1 J>_\rho\\
\hspace{1cm}\sum\limits_{\rho'}
\left[\begin{array}{cccc} (10,4) &(0,0) &(10,4) &1\\
(18,0) &(\lambda_0 \mu_0 ) &(18,4) &\rho' \\
(28,4) &(\lambda_0 \mu_0 ) &(\lambda \mu )_\sigma &\rho \\
1 &1 &1 \end{array} \right]
<(18,4)\mid\mid\mid [a^\dagger_{\tilde \eta 0),{1\over 2}}
a^\dagger_{\tilde \eta 0),{1\over 2}}]^{(\lambda_0 \mu_0 )}
\mid\mid\mid (18,0)>_{\rho'} \label{mgt}
\end{array}
\end{equation}

In the above formula $<..,..\|,,>$ denotes the SU(3) Clebsch-Gordan
coefficients \cite{Dra73}, the symbol  $[...]$ represents 
$9-\lambda\mu$ recoupling coefficients \cite{Mil78},
$<..\mid\mid\mid ..\mid\mid\mid ..>$ are the triple reduced matrix
elements \cite{Hir95c} and the following notation was introduced:

\begin{equation}
\begin{array}{l}
a = {{4\eta}\over{(2\eta + 1)\sqrt{2\eta - 1}}},
\hspace{4cm}
b(n^A_\pi ) = [n^A_\pi(\eta +2 -
n^A_\pi/2)]^{1/2},\\ ~\\
            {\cal E}_{\sigma} =
E_{\sigma} -\hbar \omega k_\pi 2 j_\pi + \Delta_C
\hspace{.4cm}\hbox{ with  } \hspace{.4cm}
\Delta_C ={ 0.70 \over A^{1/3}} [2 Z + 1 - 0.76 ( (Z+1)^{4/3} -Z^{4/3} )]
\end{array}
\end{equation}
The SU(3) tensorial components $(\lambda_0 ,\mu_0)$ of the normal part
of the double Gamow-Teller operator must be able to couple the proton and
total irreps (18,0) and (28,4) associated with the ground state of
$^{156}$Gd to the corresponding irreps (18,4) and $(\lambda
\mu)_\sigma $ = (28,8), (30,4) and (32,0), which
characterize the ground and excited rotational bands in $^{156}$Gd. If
these
irreps cannot be connected by $(\lambda_0 ,\mu_0 )$
the $\beta^+ \beta^+$ ~decays to the $0^+$ states are
forbidden. This is a
second selection rule imposed by the model to the $\beta \beta$ ~decay.

\begin{center}
\begin{tabular}{llccr}
\hline
\multicolumn{2}{c}{Transition} & $G_{2\nu}$(yr$^{-1}$) & $| \
M_{2\nu}(\sigma) \ |$ & $\tau^{1/2}$(yr) \\ \hline
$^{156}$Dy $\rightarrow ^{156}$Gd & $0^{+} \rightarrow 0^{+}(g.s.)$
& 9.79$\times 10^{-21}$ & 0.061 & 2.74$\times 10^{22}$ \\
& $0^{+} \rightarrow 0^{+}(1)$ & 1.65$\times 10^{-22}$
&0.027 &
8.31$\times 10^{24}$ \\
& $0^{+} \rightarrow 0^{+}(2)$ & 7.58$\times 10^{-23}$ &
0.035 &
1.08 $\times 10^{25}$ \\ \hline
$^{162}$Er $\rightarrow ^{162}$Dy & $0^{+} \rightarrow 0^{+}(g.s.)$
 & 8.06$\times 10^{-21}$ & 0.066   & 2.85$\times10^{22}$
\\
& $0^{+} \rightarrow 0^{+}(1)$ & 1.60$\times 10^{-24}$
& 0.013 & 3.70$\times10^{27}$ \\ \hline
$^{168}$Yb $\rightarrow ^{168}$Er & $0^{+} \rightarrow 0^{+}(g.s.)$ &
 2.47$\times 10^{-21}$ & 0.045 & 2.00$\times 10^{23}$ \\
& $0^{+} \rightarrow 0^{+}(1)$ &  5.18$\times 10^{-28}$
&0.0006 & 5.36$
\times 10^{33}$ \\ \hline
\end{tabular}
\end{center}
Table 1: Half-lives for the $ECEC_{2\nu}$ decay to the ground and excited
states of the final nuclei.\\

In Table 1 we present the $ECEC_{2\nu}$ decay of $^{156}$Dy, $^{162}$Er
and $^{168}$Yb to the ground and excited states of $^{156}$Gd, $^{162}$Dy
and $^{168}$Er respectively. The kinematical factors $G_{2\nu}(\sigma)$
were evaluated following the prescriptions given in \cite{Doi88}. 
When the energy released in the decay to an excited stated ($2 E_\sigma$)
is small the available phase space $G_{2\nu}$ is strongly reduced, and
 the half life could be very large. It is the case in the
double electron capture decay to the first excited $0^+$ state in
$^{168}$Er. 
Details
of the calculation of the nuclear matrix elements, as well as the energy
spectra of the nuclei involved will be presented elsewhere \cite{Cer99}. 
The nuclear matrix elements associated with the decay to the ground state
of the final nuclei $| M_{2\nu}(g.s.) \ |$ have values close to 0.05 -
0.06, a factor of 5 smaller than the assumption of Barabash \cite{Bar94},
and are similar for the three nuclei studied.  The nuclear matrix elements
to excited states $| M_{2\nu}(1,2) \ |$ show a wide spread, being
close to those of the ground state for $^{156}$Dy, suppressed by a factor
5 for $^{162}$Er and by a factor 80 for $^{168}$Yb. While in general it is
confirmed that  deformed nuclei have smaller nuclear matrix elements than
spherical \cite{Cas94}, $^{156}$Dy appears to be the best candidate of
this group for experimental detection, with a half-life around $10^{24}$
years for the double electron capture to the first excited $0^+$ state.

In summary, we have used the pseudo SU(3) shell model to investigate the
double electron capture decays in three heavy deformed nuclei, and found
that experiments with sensitivities around $10^{24}$ years could detect the
decay of $^{156}$Dy to the first excited state in $^{156}$Gd, while
detecting the doubled electron capture decay in $^{162}$Er and $^{168}$Yb
would be very difficult. 

The authors thanks Esteban Roulet and Roelof Bijker for the critical
reading of the manuscript. Partial economical support by Conacyt,
M\'exico, is acknowledged.

Figure Captions:

Energy levels relevant for the $\beta^+\beta^+$ decay of $^{156}$Dy.

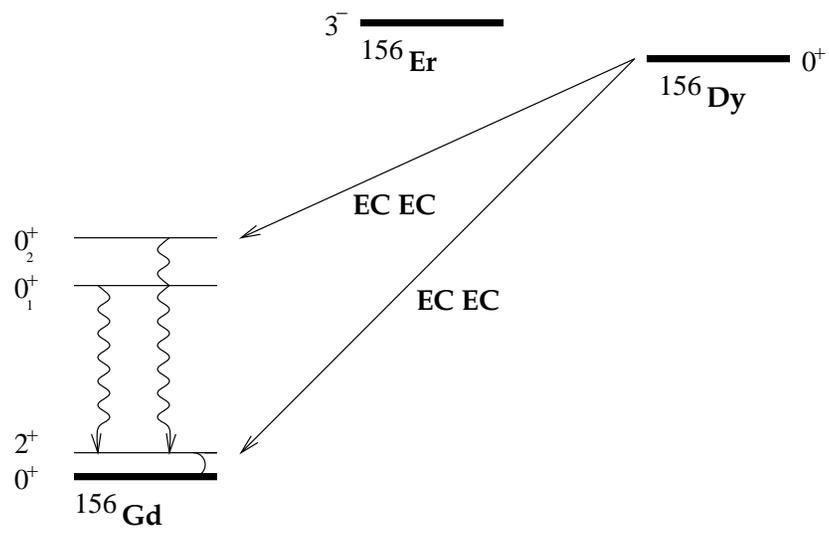
\end{document}